\documentclass[11pt,epsf,letterpaper]{article}%
\usepackage{setspace}
\usepackage{color}
\usepackage{amsmath}
\usepackage{amsfonts}
\usepackage{subcaption}
\usepackage{comment}
\usepackage{booktabs}
\usepackage{verbatim}
\usepackage{amssymb}
\usepackage{graphicx}%
\providecommand{\U}[1]{\protect\rule{.1in}{.1in}}
\begin{document}

\title{Ordered arrays of Baryonic tubes in the Skyrme model in (3+1) dimensions at
finite density}
\author{Fabrizio Canfora$^{1}$\\$^{1}$\textit{Centro de Estudios Cient\'{\i}ficos (CECS), Casilla 1469,
Valdivia, Chile.}\\{\small canfora@cecs.cl}}
\maketitle

\begin{abstract}
A consistent ansatz for the Skyrme model in (3+1)-dimensions which is able to
reduce the complete set of Skyrme field equations to just one equation for the
profile in situations in which the Baryon charge can be arbitrary large is
introduced: moreover, the field equation for the profile can be solved
explicitly. Such configurations describe ordered arrays of Baryonic tubes
living in flat space-times at finite density. The plots of the energy density
(as well as of the Baryon density) clearly show that the regions of maximal
energy density have the shape of a tube: the energy density and the Baryon
density depend periodically on two spatial directions while they are constant
in the third spatial direction. Thus, these topologically non-trivial
crystal-like solutions can be intepreted as configurations in which most of
the energy density and the baryon density are concentrated within tube-shaped
regions. The positions of the energy-density peaks can be computed explicitly
and they manifest a clear crystalline order. A non-trivial stability test is discussed.

\end{abstract}

\section{Introduction}

A theoretical description of cold and dense nuclear matter as a function of
baryon number density \cite{R1} \cite{R2} is still lacking. Analytic solutions
in the low-energy limit of QCD able to describe crystals-like structures in
Baryonic matter at finite density would be extremely helpful in this respect.
Until quite recently, the only theoretical arguments supporting the appearance
of Baryonic crystals at finite density could be found in two-dimensional
effective toy models (see \cite{toy1}, \cite{toy2}\ and references therein) as
well as in the case of the so-called magnetic Skyrmions \cite{aprox0}. Clever
approximations related to either the so-called product ansatz or to Yang-Mills
instantons may lead to analytic results in (3+1)-dimensions \cite{aproxAM}
\cite{aprox1} \cite{aprox2} \cite{aprox3} \cite{aprox4} \cite{aprox4.1}
\cite{aprox5} \cite{aprox5.1} \cite{aprox5.2} which help to clarify many
properties of these multi-solitonic configurations\footnote{Even if the
Skyrmions can be well approximated by the holonomy of the Yang--Mills
instantons, the natural limit of low-energy QCD corresponds to the instanton
liquid rather than a crystal. Thus, the analogy between Skyrmions and holonomy
of instantons is not complete: there are plenty of results supporting the
appearance of Skyrme crystals. I thank the anonimous referee for this
remark.}. In fact, it is still an open problem to build crystals of Baryons
explicitly in the low-energy limit of QCD. For this reason, different schemes
to describe this phase have been introduced.

A common approach (based on the AdS/CFT correspondence \cite{malda1}) is in
the references \cite{sakai} \cite{pop1} \cite{pop1.1} \cite{pop2} \cite{pop3}.
The main difficulty in these cases is to calculate the required solitons: only
single baryon solutions have been discussed in details \cite{critic1}
\cite{critic2}. Numerical solutions at finite density in full Sakai-Sugimoto
model are extremely difficult (see \cite{diffreview} \cite{diffreview2}).

Despite that, it has been argued that baryons in such a regime \textit{are
necessarily in a solid crystalline phase} \cite{pop2} \cite{BoSut}
\cite{Elliot} \cite{Elliot-Ripley:2016ctk}. More complex ordered structures
are also expected to appear at finite densities. It is widely accepted that
the so-called \textit{nuclear pasta phase} do exist (see \cite{pasta0}
\cite{pasta01} \cite{pasta02} \cite{pasta} and references therein). In this
phase, the energy-density plots in the references just mentioned show, for
instance, that ordered configurations of "Baryonic tubes" (\textit{nuclear
spaghetti}) in which most of the Baryonic charge is gathered in tube-shaped
regions are observed (\textit{nuclear lasagna} and \textit{nuclear gnocchi}
phases have also been observed). It is usually assumed that the available
theoretical tools are unsuitable to describe such complex structures in the
low energy limit of (only numerical results are available in the nuclear pasta phase).

An explicit analytic approach to describe the predictions above in the context
of the low energy limit of (3+1)-dimensional QCD is the main goal of the
present paper.

The Skyrme theory represents such low energy limit \cite{skyrme} (for detailed
reviews on Skyrme theory see \cite{multis2}\ \cite{manton}). Its topological
solitons (called Skyrmions) represent Baryons (see \cite{bala0} \cite{witten0}
\cite{ANW} and references therein). The wide range of applications of this
theory in many different fields (see \cite{useful2}, \cite{useful3},
\cite{useful6}, \cite{useful7.1}, \cite{useful7} and references therein) is
well recognized.

The appearance of crystal-like structures in the Skyrme model (which was
predicted on the basis of the product ansatz introduced in \cite{aprox1}) is
well established numerically (see \cite{manton}\ \cite{rational} and
references therein). Within the rational map approach \cite{rational} one can
construct numerically configurations in which the number of "bumps" in the
energy density is related with the corresponding Baryon charge. Based on the
fact that, in the Skyrme model, the solution representing one spherical
Skyrmion (introduced by Skyrme himself) must be found numerically, it has been
always assumed that it is impossible to find analytic solutions representing
ordered pattern of solitons without approximations. Consequently, interesting
modifications of the original Skyrme model have been considered which do allow
to find analytic solutions representing Skyrmions at finite Baryon density
(see, in particular, \cite{aprox7} \cite{aprox8} \cite{aprox9} \cite{aprox10}%
\ and references therein). However, only the original Skyrme model (which is
directly related with the low energy limit of QCD) will be considered here.

The generalized hedgehog ansatz introduced in \cite{56} \cite{56b} \cite{58}
\cite{58b}\ \cite{ACZ} \cite{CanTalSk1} \cite{Fab1} \cite{gaugsk}
\cite{gaugsk2}\ allowed the construction of the first analytic multi-Skyrmions
at finite density (however, these Skyrmions at finite density do not have
crystal-like structure).

In the present paper, the methods in \cite{Fab1} \cite{gaugsk} \cite{gaugsk2}
will be generalized to construct ordered arrays of Baryonic tubes with
crystalline structure living at finite density. One then observes transitions
in which for fixed total topological charge, configurations with taller tubes
but a smaller number of tubes are energetically favored over configurations
made of a lower tubes but with a higher number of tubes. 

This paper is organized as follows: in the second section the Skyrme model is
introduced and the Skyrme field equations are written in two equivalent ways:
this is very important in order to check that the present crystals are really
solutions of the full Skyrme field equations. In the third section, the method
to go beyond the spherical hedgehog ansatz is introduced and the analytic
crystal-like Skyrmions are derived. In the fourth section the appearance of
fractional Baryonic charge is discussed. In the final section, some
conclusions will be drawn.

\section{Beyond the spherical hedgehog ansatz}

The action of the $SU(2)$ Skyrme system is
\begin{align}
S_{Sk}  &  =\frac{K}{2}\int d^{4}x\sqrt{-g}\,\mathrm{Tr}\left(  \frac{1}%
{2}L^{\mu}L_{\mu}+\frac{\lambda}{16}F_{\mu\nu}F^{\mu\nu}\right)
\ \ \ K>0\ ,\ \ \ \lambda>0\ ,\label{skyrmaction}\\
L_{\mu}  &  :=U^{-1}\nabla_{\mu}U=L_{\mu}^{j}t_{j}\ ,\ \ F_{\mu\nu}:=\left[
L_{\mu},L_{\nu}\right]  \ ,\ \ \ \hbar=1\ ,\ \ \ c=1\ ,\nonumber
\end{align}
where $K$ and $\lambda$ are the coupling constants\footnote{They can be
determined as in \cite{ANW}.}, $\mathbf{1}_{2}$\ is the $2\times2$ identity
matrix and the $t^{j}$ are the basis of the $SU(2)$ generators (where the
Latin index $j$ corresponds to the group index).

The three coupled Skyrme field equations can be obtained by taking the
variation of the action with respect to the $SU(2)$-valued field $U$:
\begin{equation}
\frac{\delta S_{Sk}}{\delta U}=0\Leftrightarrow\nabla^{\mu}L_{\mu}%
+\frac{\lambda}{4}\nabla^{\mu}[L^{\nu},F_{\mu\nu}]=E^{j}t_{j}=0\ ,\ j=1,2,3.
\label{nonlinearsigma1}%
\end{equation}

The energy density (the $0-0$ component of the energy-momentum tensor) reads
\begin{equation}
T_{00}=-\frac{K}{2}\mathrm{Tr}\left[  L_{0}L_{0}-\frac{1}{2}g_{00}L^{\alpha
}L_{\alpha}\,+\frac{\lambda}{4}\left(  g^{\alpha\beta}F_{0\alpha}F_{0\beta
}-\frac{g_{00}}{4}F_{\sigma\rho}F^{\sigma\rho}\right)  \right]  \ .
\label{timunu1}%
\end{equation}

The following parametrization of the $SU(2)$-valued scalar $U(x^{\mu})$ will
be adopted%
\begin{align}
U^{\pm1}(x^{\mu})  &  =\cos\left(  \alpha\right)  \mathbf{1}_{2}\pm\sin\left(
\alpha\right)  n^{i}t_{i}\ ,\ \ n^{i}n_{i}=1\ ,\label{standard1}\\
n^{1}  &  =\sin F\cos G\ ,\ \ n^{2}=\sin F\sin G\ ,\ \ n^{3}=\cos
F\ ,\label{standard1.1}\\
\alpha &  =\alpha\left(  x^{\mu}\right)  \ ,\ F=F\left(  x^{\mu}\right)
\ ,\ G=G\left(  x^{\mu}\right)  \label{standard1.112}%
\end{align}
where, generically, $\alpha$, $F$\ and $G$\ are functions of the four
space-time coordinates.

It is worth to emphasize here a relevant technical point. The number of Skyrme
field equations is equal to the dimension of the corresponding Lie Algebra (in
the present case the $SU(2)$ case will be considered so that the task is to
solve three coupled non-linear field equations). One can write down the
complete set of Skyrme field equations in many different equivalent ways. In
this paper two natural representations of the field equations will be
considered. The first one is described in Eq. (\ref{nonlinearsigma1}). In this
matrix representation the three field equations (namely, $E^{j}=0$,\ $j=1$,
$2$, $3$) are obtained as variation of the action with respect to the
$SU(2)$-valued scalar field $U$. An equivalent way (which will be described in
the next subsection) corresponds to take the most general representation of
$SU(2)$-valued scalar field $U$ in the terms of three scalar degrees of
freedom ($\alpha$, $F$ and $G$ defined in Eqs. (\ref{standard1}),
(\ref{standard1.1})) and (\ref{standard1.112})) and replace such general
representation of $U$ into the action (which will become an action for the
three interacting scalar fields $\alpha$, $F$ and $G$). In this second
representation, the three field equations (which will be derived explicitly in
the next subsection) will be simply%
\[
\frac{\delta S_{Sk}}{\delta\alpha}=0\ ,\ \frac{\delta S_{Sk}}{\delta
F}=0\ ,\ \frac{\delta S_{Sk}}{\delta G}=0\ .
\]
Obviously, the two representations are equivalent (in particular, it will be
shown explicitly in both representations that \textit{the analytic crystals
constructed in the present paper are solutions of the complete set of Skyrme
field equations}). The only warning is that, in general, the field equations
in one representation are linear combinations of the field equations in the
second representation.

The most difficult task is to find a good ansatz which keeps alive the
non-trivial topological charge and, at the same time, allows for a
crystal-like structure in the energy density without making the field
equations impossible to solve analytically. It is worth to note that with the
usual hedgehog ansatz in \cite{skyrme} (as well as in its finite density
generalization \cite{Fab1} \cite{gaugsk}) the energy density in Eq.
(\ref{timunu1}) (due to the trace on the internal indices) only depends on the
Skyrmion profile. Until now, this fact prevented one from describing
analytically multi-solitonic configurations with crystal-like order. It is
also interesting to note that the main building blocks of these ordered
multi-solitonic structures are not necessarily single nucleons. Indeed, it is
widely accepted that the so-called \textit{nuclear pasta phase} may appear
(see \cite{pasta0} \cite{pasta01} \cite{pasta02} \cite{pasta} and references
therein). In this phase one can have, for instance, ordered arrays of
"Baryonic tubes" (\textit{nuclear spaghetti}) in which the shapes of the
regions of maximal energy density and Baryon density are tubes. To describe
such configurations one necessarily needs an ansatz in which both the Baryon
density and the energy density depend not only on the profile $\alpha$ but
also on an extra spatial coordinate.

The Baryon charge of the configuration reads%
\begin{equation}
W=B=\frac{1}{24\pi^{2}}\int_{\left\{  t=const\right\}  }\rho_{B}\ ,
\label{rational4}%
\end{equation}%
\begin{equation}
\rho_{B}=\epsilon^{ijk}Tr\left(  U^{-1}\partial_{i}U\right)  \left(
U^{-1}\partial_{j}U\right)  \left(  U^{-1}\partial_{k}U\right)  \ .
\label{rational4.1}%
\end{equation}

In terms of $\alpha$, $F$ and $G$,\ the topological density $\rho_{B}$ reads
\begin{equation}
\rho_{B}=12\left(  \sin^{2}\alpha\sin F\right)  d\alpha\wedge dF\wedge
dG\ ,\label{rational4.1.1}%
\end{equation}
so that a necessary condition in order to have non-trivial topological charge
is%
\begin{equation}
d\alpha\wedge dF\wedge dG\neq0\ .\label{necscond}%
\end{equation}
From the geometrical point of view the above condition (which simply states
that $\alpha$, $F$ and $G$ must be three independent functions) can be
interpreted as saying that such three functions "fill a three-dimensional
spatial volume" at least locally. In other words, $d\alpha$, $dF$ and $dG$ can
be used as 3D volume form. Hence, the condition in Eq. (\ref{necscond})
ensures that the configuration one is interested in describes a genuine
three-dimensional object. On the other hand, such a condition is not
sufficient in general. One has also to require that the spatial integral of
$\rho_{B}$ must be a \textit{non-vanishing integer}:%
\begin{equation}
\frac{1}{24\pi^{2}}\int_{\left\{  t=const\right\}  }\rho_{B}\in%
\mathbb{Z}
\ .\label{necscond2}%
\end{equation}
Usually, this second requirement allows to fix some of the parameters of the ansatz.

From now on, as it is customary in the literature, the terms
\textit{Skyrmionic crystals} (and also \textit{crystals of solitons}) will
refer to smooth regular solutions of Eq. (\ref{nonlinearsigma1}) with the
properties that both the topological charge (defined in Eqs. (\ref{rational4}%
), (\ref{rational4.1}), (\ref{rational4.1.1}) and (\ref{necscond})) is
non-vanishing and larger than 1 and, moreover, that the local maxima of energy
density and Baryon density manifest a crystal-like ordered structure.

\subsection{Explicit parametrization}

It is useful to write down the Skyrme field equations explicitly in terms of
the three scalar degrees of freedom $\alpha$, $F$ and $G$. In this way, it is
possible to check directly that the novel ansatz proposed in the present paper
is consistent. It is convenient to introduce the following functions
$Y^{0}(x^{\mu})$ and $Y^{i}(x^{\mu})$
\begin{align}
Y^{0}  &  =\cos\alpha,\ Y^{1}=\sin(\alpha)\sin(F)\cos
(G),\ \label{parametrization}\\
Y^{2}  &  =\sin(\alpha)\sin(F)\sin(G),\ Y^{3}=\sin(\alpha)\cos(F)\ .
\label{parametrization0}%
\end{align}
In this way, the most general element $U$ of $SU(2)$ can be written as
\[
U=\mathbf{I}Y^{0}+Y^{i}t_{i}\;\;\;;\;\;\;U^{-1}=\mathbf{I}Y^{0}-Y^{i}t_{i}\ .
\]
These functions are useful to write down the Skyrme action explicitly in terms
of $\alpha$, $F$ and $G$. Introducing the tensor $\Sigma_{\mu\nu}$
\begin{equation}
\Sigma_{\mu\nu}=G_{ij}\nabla_{\mu}Y^{i}\nabla_{\nu}Y^{j}\;\;;\;\;G_{ij}%
=\delta_{ij}+\frac{Y^{i}Y^{j}}{1-Y^{k}Y_{k}} \label{sigma1}%
\end{equation}
the Skyrme action is then defined as
\begin{equation}
I=\int d^{4}x\sqrt{-g}\left[  \frac{1}{2}\Sigma_{\mu}^{\mu}+\frac{\lambda}%
{4}\left(  (\Sigma_{\mu}^{\mu})^{2}-\Sigma^{\mu\nu}\Sigma_{\mu\nu}\right)
\right]  . \label{sigma2}%
\end{equation}

Now, we are in the position to write down the general Skyrme field equations
in terms of $\alpha$, $F$ and $G$ (which obviously are equivalent to the
Skyrme field equations in matrix form in Eq. (\ref{nonlinearsigma1})). The
variation of the Skyrme action with respect to $\alpha$ leads to the equation
of motion
\begin{equation}%
\begin{array}
[c]{c}%
\left(  -\Box\alpha+\sin(\alpha)\cos(\alpha)\left(  \nabla_{\mu}F\nabla^{\mu
}F+\sin^{2}F\nabla_{\mu}G\nabla^{\mu}G\right)  \right) \\
+\lambda\left(
\begin{array}
[c]{c}%
\sin(\alpha)\cos(\alpha)\left(  (\nabla_{\mu}\alpha\nabla^{\mu}\alpha
)(\nabla_{\nu}F\nabla^{\nu}F)-(\nabla_{\mu}\alpha\nabla^{\mu}F)^{2}\right) \\
+\sin(\alpha)\cos(\alpha)\sin^{2}(F)\left(  (\nabla_{\mu}\alpha\nabla^{\mu
}\alpha)(\nabla_{\nu}G\nabla^{\nu}G)-(\nabla_{\mu}\alpha\nabla^{\mu}%
G)^{2}\right) \\
+2\sin^{3}(\alpha)\cos(\alpha)\sin^{2}(F)\left(  (\nabla_{\mu}F\nabla^{\mu
}F)(\nabla_{\nu}G\nabla^{\nu}G)-(\nabla_{\mu}F\nabla^{\mu}G)^{2}\right) \\
-\nabla_{\mu}\left(  \sin^{2}(\alpha)(\nabla_{\nu}F\nabla^{\nu}F)\nabla^{\mu
}\alpha\right)  +\nabla_{\mu}\left(  \sin^{2}(\alpha)(\nabla_{\nu}\alpha
\nabla^{\nu}F)\nabla^{\mu}F\right) \\
-\nabla_{\mu}\left(  \sin^{2}(\alpha)\sin^{2}(F)(\nabla_{\nu}G\nabla^{\nu
}G)\nabla^{\mu}\alpha\right)  +\nabla_{\mu}\left(  \sin^{2}(\alpha)\sin
^{2}(F)(\nabla_{\nu}\alpha\nabla^{\nu}G)\nabla^{\mu}G\right)
\end{array}
\right)
\end{array}
=0\ , \label{equ}%
\end{equation}
The variation of the Skyrme action with respect to $F$ leads to the equation
of motion
\begin{equation}%
\begin{array}
[c]{c}%
\left(  -\sin^{2}(\alpha)\Box F-2\sin(\alpha)\cos(\alpha)\nabla_{\mu}%
\alpha\nabla^{\mu}F+\sin^{2}(\alpha)\sin(F)\cos(F)\nabla_{\mu}G\nabla^{\mu
}G\right) \\
+\lambda\left(
\begin{array}
[c]{c}%
\sin^{2}(\alpha)\sin(F)\cos(F)\left(  (\nabla_{\mu}\alpha\nabla^{\mu}%
\alpha)(\nabla_{\nu}G\nabla^{\nu}G)-(\nabla_{\mu}\alpha\nabla^{\mu}%
G)^{2}\right) \\
+\sin^{4}(\alpha)\sin(F)\cos(F)\left(  (\nabla_{\mu}F\nabla^{\mu}%
F)(\nabla_{\nu}G\nabla^{\nu}G)-(\nabla_{\mu}F\nabla^{\mu}G)^{2}\right) \\
-\nabla_{\mu}\left(  \sin^{2}(\alpha)(\nabla_{\nu}\alpha\nabla^{\nu}%
\alpha)\nabla^{\mu}F\right)  +\nabla_{\mu}\left(  \sin^{2}(\alpha)(\nabla
_{\nu}\alpha\nabla^{\nu}F)\nabla^{\mu}\alpha\right) \\
-\nabla_{\mu}\left(  \sin^{4}(\alpha)\sin^{2}(F)(\nabla_{\nu}G\nabla^{\nu
}G)\nabla^{\mu}F\right)  +\nabla_{\mu}\left(  \sin^{4}(\alpha)\sin
^{2}(F)(\nabla_{\nu}F\nabla^{\nu}G)\nabla^{\mu}G\right)
\end{array}
\right)
\end{array}
=0\ , \label{equ1.1}%
\end{equation}
The variation of the Skyrme action with respect to $G$ leads to the equation
of motion
\begin{equation}%
\begin{array}
[c]{c}%
\left(  -\sin^{2}(\alpha)\sin^{2}(F)\Box G-2\sin(\alpha)\cos(\alpha)\sin
^{2}(F)\nabla_{\mu}\alpha\nabla^{\mu}G-2\sin^{2}(\alpha)\sin(F)\cos
(F)\nabla_{\mu}F\nabla^{\mu}{G}\right) \\
+\lambda\left(
\begin{array}
[c]{c}%
-\nabla_{\mu}\left[  \sin^{2}(\alpha)\sin^{2}(F)(\nabla_{\nu}\alpha\nabla
^{\nu}\alpha)\nabla^{\mu}G\right]  +\nabla_{\mu}\left[  \sin^{2}(\alpha
)\sin^{2}(F)(\nabla_{\nu}\alpha\nabla^{\nu}G)\nabla^{\mu}\alpha\right] \\
-\nabla_{\mu}\left[  \sin^{4}(\alpha)\sin^{2}(F)(\nabla_{\nu}F\nabla^{\nu
}F)\nabla^{\mu}G\right]  +\nabla_{\mu}\left[  \sin^{4}(\alpha)\sin
^{2}(F)(\nabla_{\nu}F\nabla^{\nu}G)\nabla^{\mu}F\right]
\end{array}
\right)
\end{array}
=0\ . \label{equ1.2}%
\end{equation}

\subsubsection{Example: the original Skyrme ansatz}

Here we give a simple and well known example showing that the above equations
(\ref{equ}), (\ref{equ1.1}) and (\ref{equ1.2}) are suitable to devise a
strategy to find good ansatz which reduce the full Skyrme field equations to
only one consistent equation for the profile in a non-trivial topological
sector. The original Skyrme ansatz is defined by the choice%
\begin{equation}
\alpha=\alpha_{s}(R)\ ,\ F=\theta\ ,\ G=\varphi\ , \label{original1}%
\end{equation}
in the flat metric in spherical coordinates:%
\begin{equation}
ds^{2}=g_{\mu\nu}dx^{\mu}dx^{\nu}=-dt^{2}+dR^{2}+R^{2}\left(  d\theta^{2}%
+\sin\theta^{2}d\varphi^{2}\right)  \ . \label{original2}%
\end{equation}
One can check that, due to the identities%
\begin{equation}
\nabla_{\mu}\alpha_{s}\nabla^{\mu}F=\nabla_{\mu}G\nabla^{\mu}\alpha_{s}%
=\nabla_{\mu}G\nabla^{\mu}F\ =0\ , \label{original3}%
\end{equation}
and to the fact that both $F$ and $G$ are linear functions in the chosen
coordinates system, the three Skyrme field equations (\ref{equ}),
(\ref{equ1.1}) and (\ref{equ1.2}) reduce to only one consistent scalar
equation for the profile $\alpha_{s}(R)$. \textit{A crucial technical detail
is the following}: in the equation for $\alpha_{s}(R)$ (namely, Eq.
(\ref{equ})) potentially dangerous terms are the ones involving $\sin F^{2}$
as these terms involve $\theta$ while one would like to have a consistent
equation for $\alpha_{s}(R)$ which, therefore, can only involve $R$%
-dependence. In the original ansatz of Skyrme the dangerous $\theta
-$dependence due to $\sin F^{2}$ is canceled by the inverse metric
$g^{\varphi\varphi}$ appearing in $\nabla_{\mu}G\nabla^{\mu}G$. Thus, one can
reduce the three Skyrme field equations to a single scalar equation for
$\alpha_{s}(R)$. If one plugs the ansatz in Eq. (\ref{original1}) into the
three Skyrme field equations (\ref{equ}), (\ref{equ1.1}) and (\ref{equ1.2})
one can see directly that Eqs. (\ref{equ1.1}) and (\ref{equ1.2}) are
identically satisfied and that Eq. (\ref{equ}) reduces to the usual equation
for the Skyrme profile $\alpha_{s}(R)$ which can be found in all the textbooks
(see, for instance, \cite{manton}). In the following, a different trick to get
a consistent equation for $\alpha$ in a topologically non-trivial sector will
be introduced.

\subsubsection{Comparison with the rational map ansatz}

The idea of the rational map approach \cite{rational}\ is to obtain an
approximate description of multi-Skyrmionic configurations replacing the
isospin vector $n^{i}$ for the spherical hedgehog defined in Eqs.
(\ref{standard1.1}), (\ref{original1}) and (\ref{original2}) by a more general
map between two spheres. Even if the rational map ansatz is only an
approximation, it provides one with a very clear intuitive picture of
multi-Skyrmionic configurations.

The usual starting point is%
\begin{equation}
Y^{0}=\cos\alpha\ ,\ \ Y^{i}=\left(  n_{\Psi}\right)  ^{i}\sin\alpha
\ ,\ \ \ \alpha=\alpha(R)\ ,\ \ \delta_{ij}\left(  n_{\Psi}\right)
^{i}\left(  n_{\Psi}\right)  ^{j}=1\label{rational1}%
\end{equation}%
\begin{align}
\left(  n_{\Psi}\right)  ^{1} &  =\frac{\Psi+\overline{\Psi}}{1+\left\vert
\Psi\right\vert ^{2}}\ ,\ \ \ \left(  n_{\Psi}\right)  ^{2}=\frac{i\left(
\Psi-\overline{\Psi}\right)  }{1+\left\vert \Psi\right\vert ^{2}%
}\ ,\ \ \ \left(  n_{\Psi}\right)  ^{3}=\frac{1-\left\vert \Psi\right\vert
^{2}}{1+\left\vert \Psi\right\vert ^{2}}\ ,\label{rational2}\\
\Psi &  =\Psi(z)\ ,\ \ z\in%
\mathbb{C}
\ \ ,\label{rational3}%
\end{align}
where $z$ is a complex coordinate which, using the stereographic projection,
can be identified with the coordinates on the 2-sphere:%
\[
z=\exp\left(  i\varphi\right)  \tan\left(  \frac{\theta}{2}\right)  \ .
\]
The rational map $\Psi(z)$ (which gives the name to the whole approach) reads%
\[
\Psi\left(  z\right)  =\frac{p\left(  z\right)  }{q\left(  z\right)  }\ ,
\]
where $p$ and $q$ are polynomials in $z$ with no common factor. The degree $N$
of the rational map $\Psi$ is defined as%
\begin{equation}
N=\int\frac{2idzd\overline{z}}{\left(  1+\left\vert z\right\vert ^{2}\right)
^{2}}\left(  \frac{1+\left\vert z\right\vert ^{2}}{1+\left\vert \Psi
\right\vert ^{2}}\left\vert \frac{d\Psi}{dz}\right\vert \right)
^{2}\ .\label{rational3.5}%
\end{equation}
It is well-known that%
\[
N=\max\left(  n_{p},n_{q}\right)  \ ,
\]
where $n_{p}$ and $n_{q}$ are the degrees of $p\left(  z\right)  $ and
$q\left(  z\right)  $ respectively. Correspondingly, the Baryon number $B$ of
the configuration in Eqs. (\ref{rational1}), (\ref{rational2}) and
(\ref{rational3}) reads%
\begin{equation}
B=-\frac{1}{24\pi^{2}}\int\epsilon^{ijk}Tr\left(  U^{-1}\partial_{i}U\right)
\left(  U^{-1}\partial_{j}U\right)  \left(  U^{-1}\partial_{k}U\right)
=nN\ ,\label{rational44}%
\end{equation}
where%
\begin{equation}
n=\left(  -\frac{2}{\pi}\int\left(  \alpha^{\prime}\sin^{2}\alpha\right)
dR\right)  \ .\label{rational44.1}%
\end{equation}
The winding number is the product of the contribution coming from the profile
function $\alpha$ times the degree $N$ of the rational map $\Psi$. One can
think that $n$ is the number of "bumps" associated to $\alpha$ while $N$ are
related to the bumps in the directions orthogonal to $\alpha$.

When (and only when)%
\begin{equation}
\Psi=z\ , \label{standard}%
\end{equation}
the rational map ansatz in Eqs. (\ref{rational1}), (\ref{rational2}) and
(\ref{rational3}) reduces to the original spherical hedgehog ansatz discussed
in the previous subsection and it gives rise to a (numerical) solution of the
complete set of Skyrme field equations. On the other hand, in all the higher
degrees cases $N>1$ the rational map solutions are only approximated solutions
of the complete set of Skyrme field equations.

The strategy\footnote{In other words, one first minimizes the energy
functional with respect to the rational map (with fixed degree $N$). Then, one
is left with an energy functional which only depends on the profile function
so that the minimization procedure is reduced to a one-dimensional problem.}
of this approach is to minimize the total energy with respect to both $\alpha$
and the rational map $\Psi$. This framework is tied to situations in which
there is an approximated spherical symmetry: in particular, the crystals-like
configurations typical of nuclear pasta (see \cite{pasta0} \cite{pasta01}
\cite{pasta02} \cite{pasta} and references therein) cannot be analyzed in this
approach. The main advantage of this rational-map framework is that it
disentangles the radial coordinate $R$ from the angular coordinates. The main
disadvantage is that, as already emphasized, such a disentangling is only an approximation.

In terms of the explicit parametrization introduced in Eqs.
(\ref{parametrization}) and (\ref{parametrization0}) (in which the complete
set of Skyrme field equations are (\ref{equ}), (\ref{equ1.1}) and
(\ref{equ1.2})) one can interpret the rational map approach as follows. The
first step of the procedure (in which one minimizes the energy functional with
respect to the rational map) gives rise to an effective equation for the
radial profile which is very similar to the original one of Skyrme with the
difference that some parameters are rescaled (see, in particular, the
discussion in \cite{CanTalSk1} and references therein). Thus, in a sense, one
can think that in the rational-map approach Eqs. (\ref{equ1.1}) and
(\ref{equ1.2}) are solved "on average" (freezing $\alpha$) and then the
corresponding Eq. (\ref{equ}) for $\alpha$ is solved (more often than not) numerically.

Instead, in the present paper it will be introduced an ansatz with three main advantages.

\textit{Firstly}, the complete set of Skyrme field equations \textit{will be
solved exactly} (and not only approximately). In other words, the ansatz
described in the next sections will solve Eqs. (\ref{equ}), (\ref{equ1.1}) and
(\ref{equ1.2}) without any approximation.

\textit{Secondly}, the solutions will be analytical and not numerical.

\textit{Thirdly}, the ansatz is not tied to situations with spherical symmetry.

\section{Skyrme Crystals}

The main physical motivation of the present work\textbf{ }is to study finite
density effects. The easiest way to take into account finite-density effects
is to introduce the following flat metric
\begin{equation}
ds^{2}=-dt^{2}+A\left(  dr^{2}+d\theta^{2}\right)  +L^{2}d\phi^{2}%
\ ,\label{Minkowski}%
\end{equation}
where $4\pi^{3}LA$ is the volume of the box in which the Skyrmions
configurations to be constructed below will live. The adimensional spatial
coordinates $r$, $\theta$ and $\phi$ have the range
\begin{equation}
0\leq r\leq2\pi\ ,\quad0\leq\theta\leq\pi\ ,\quad0\leq\phi\leq2\pi
\ .\label{period0}%
\end{equation}
Following the strategy of \cite{Fab1} \cite{gaugsk}, the boundary conditions
in the $\theta$ direction here will be chosen to be Dirichlet while in the $r$
and $\phi$\ directions they can be both periodic and anti-periodic. These are
not the only allowed boundary conditions: however this is the simplest choice
as it will be discussed in a moment.

The ansatz for the Skyrme crystal is
\begin{equation}
\alpha=\alpha\left(  r\right)  \ ,\ F=q\theta\ ,\ G=p\left(  \frac{t}{L}%
-\phi\right)  \ ,\ q=2v+1,\quad p,\ v\in%
\mathbb{N}
\ ,\ p\neq0\ . \label{ans1}%
\end{equation}
The periodic time dependence\footnote{Note that the Skyrme field $U$ depends
on the function $G$ only through $\sin G$ and $\cos G$ so that $U$ is periodic
in time.} in Eq. (\ref{ans1}) is natural for three very good reasons.

\textit{Firstly}, the most popular way to avoid Derrick's famous no-go theorem
on the existence of solitons in non-linear scalar field theories corresponds
to search for a time-periodic ansatz such that the energy density of the
configuration is still static, as it happens for boson stars \cite{5} (in the
simpler case of $U(1)$-charged scalar field: see \cite{6} and references
therein). The present ansatz defined in Eqs. (\ref{standard1}),
(\ref{standard1.1}) and (\ref{ans1}) has exactly this property. Indeed, the
energy density does not depend on time (see Eqs. (\ref{enden1}),
(\ref{enden2}) and (\ref{enden3}) below) so that these configurations possess
a static energy density.

\textit{Secondly}, unlike what happens for the usual Bosons star ansatz for
$U(1)$-charged scalar fields, the present ansatz for $SU(2)$-valued scalar
field also possesses a non-trivial topological charge.

\textit{Thirdly}, with the above choice of the ansatz the complete set of
Skyrme field equations simplifies dramatically (as it will be shown below) due
to the fact that, with the function $G$ in Eq. (\ref{ans1}), $\nabla_{\mu
}G\nabla^{\mu}G=0$.

With the ansatz defined in Eqs. (\ref{standard1}), (\ref{standard1.1}) and
(\ref{ans1}) the topological density in Eq. (\ref{rational4.1}) reads%
\begin{equation}
\rho_{B}=\left[  \left(  12pq\sin\left(  q\theta\right)  \sin^{2}%
\alpha\right)  \partial_{r}\alpha\right]  dr\wedge d\theta\wedge d\phi\ .
\label{ansch}%
\end{equation}
The above ansatz satisfies the first condition to be topologically non-trivial
in Eq. (\ref{necscond}) as the three function $\alpha$, $F$ and $G$ in Eq.
(\ref{ans1}) are independent and "fill a three-dimensional space" (see the
discussion about Eq. (\ref{necscond})). As far as the boundary conditions are
concerned, one can check directly that if one chooses, for instance, periodic
boundary conditions in $\theta$ (taking $q$ to be an even number in Eq.
(\ref{ans1})) the topological charge vanishes (even if the topological density
does not) so that the second condition in Eq. (\ref{necscond2}) would be
violated. Hence, the simplest choice to get topologically non-trivial
configurations is to take $q$ to be an odd integer. I hope to come back on the
analysis of other possible boundary conditions in a future publication. The
boundary condition on $\alpha$ and the corresponding Baryon charge are%
\begin{equation}
\alpha\left(  2\pi\right)  -\alpha\left(  0\right)  =n\pi\ \Rightarrow
\ B=np\ . \label{bc1}%
\end{equation}
Thus, the present ansatz cannot be deformed continuously to the trivial vacuum
$U_{0}=\mathbf{1}_{2x2}$ since the third homotopy class (which represents the
Baryon charge) is non-vanishing.

\subsection{Tube-shaped regions}

In order to understand the shape of the regions of maximal Baryon density
(which will coincide with the shape of the regions of maximal energy density:
see Eqs. (\ref{enden1}), (\ref{enden2}) and (\ref{enden3})\ below) one can
proceed as follows. The baryon density $\rho_{B}$ in Eqs. (\ref{rational4.1})
and (\ref{ansch}) is periodic in $r$ and $\theta$ while it is constant in the
$\phi$-direction\footnote{The fact that the Baryon density is periodic both in
$r$ and in $\theta$\ can be seen taking into account the explicit solution for
$\alpha$ in Eqs. (\ref{equ2.0}), (\ref{equ3}) and (\ref{equ3.1}). The same is
true for the energy-density: see Eqs. (\ref{enden1}), (\ref{enden2}) and
(\ref{enden3}) below.}. One observes that the positions of the maxima of the
Baryon density are defined by the conditions%
\begin{equation}
\sin^{2}\alpha=1\ ,\ \sin^{2}\left(  q\theta\right)  =1\ , \label{maxcond}%
\end{equation}
which define a periodic 2-dimensional lattice in $r$ and $\theta$. The
contour-plots in \textit{Figure 1} for the energy-density have a similar
pattern. Since both the Baryon density and the energy density are constant in
the third spatial direction $\phi$, a three-dimensional contour
plot\footnote{A three-dimensional contour plot can be easily done since all
the present solutions are known in explicit analytic form. However, I have
been unable to reduce such 3D contour plots to a reasonable size (less than 10
MB). On the other hand, the two-dimensional contour plots should be enough to
understand the shapes of the regions of maximal energy density.} would show
that the shapes of the regions of maximal energy density (and similarly for
the Baryon density) are tubes of length $2\pi L$.

In other words, in order to understand the three-dimensional shape of the
regions of maximal energy and Baryon density one has simply to take the
clear-yellow spots in \textit{Figure 1} and move them along the $\phi
$-direction (which is orthogonal to the $r-\theta$ plane of \textit{Figure
1}). Hence, such regions are (ordered arrays of) parallel tubes. The "$\phi
$-constant" sections of these tubes correspond to the contour plots in
\textit{Figure 1. }

\subsection{The explicit solution}

When one plugs the ansatz in Eq. (\ref{ans1}) into the complete set of three
coupled Skyrme field equations in Eq. (\ref{nonlinearsigma1}), they reduce to
only one integrable equation for $\alpha(r)$:%
\begin{equation}
E_{j}=c_{j}P\left[  \alpha\right]  \ ,\ c_{j}\neq0\ ,\ j=1,2,3, \label{equ1}%
\end{equation}%
\begin{equation}
P\left[  \alpha\right]  =0\Leftrightarrow\partial_{r}\left[  Y\left(
\alpha\right)  \frac{\left(  \partial_{r}\alpha\right)  ^{2}}{2}%
-V(\alpha)-E_{0}\right]  =0\ , \label{equ2}%
\end{equation}%
\begin{equation}
Y\left(  \alpha\right)  =\left(  A+q^{2}\lambda\sin^{2}\alpha\right)
\ ,\ V(\alpha)=\frac{q^{2}A}{2}\sin^{2}\alpha\ \Rightarrow\label{equ2.0}%
\end{equation}%
\begin{align}
\eta\left(  \alpha,E_{0}\right)  \overset{def}{=}\frac{\left[  2\left(
E_{0}+V(\alpha)\right)  \right]  ^{1/2}}{Y\left(  \alpha\right)  ^{1/2}}\  &
\Rightarrow\label{equ3}\\
\ \frac{d\alpha}{\eta\left(  \alpha,E_{0}\right)  }  &  =dr\ , \label{equ3.1}%
\end{align}
where $E_{0}$ is an integration constant to be fixed requiring the boundary
condition\footnote{The positive sign of the square root of $\left(
\partial_{r}\alpha\right)  ^{2}$ has been chosen.} in Eq. (\ref{bc1}). In
other words, with the ansatz in Eq. (\ref{ans1}) all the three Skyrme field
equations in Eq. (\ref{nonlinearsigma1})\ $E_{j}=0$ have the common factor
$P\left[  \alpha\right]  $ defined in Eq. (\ref{equ2}) so that if $P\left[
\alpha\right]  =0$ then \textit{all the Skyrme field equations are satisfied}.
This is the reason why the ansatz in Eq. (\ref{ans1}) has been chosen.

An alternative way to see that the above ansatz reduce the complete set of
Skyrme field equations to just one equation for the profile $\alpha$ keeping
alive the topological charge is to use the explicit parametrization of the
Skyrme action in term of $\alpha$, $F$ and $G$ in Eqs. (\ref{parametrization}%
), (\ref{sigma1}) and (\ref{sigma2}). In this way, one can read directly the
field equations for the three functions $\alpha$, $F$ and $G$. The relevant
properties of the ansatz in Eq. (\ref{ans1}) are%
\begin{align}
\nabla_{\mu}\alpha\nabla^{\mu}F  &  =\nabla_{\mu}\alpha\nabla^{\mu}%
G=\nabla_{\mu}G\nabla^{\mu}F=0\ ,\label{prop1.1}\\
\nabla_{\mu}G\nabla^{\mu}G  &  =0\ ,\ \label{prop1.2}\\
\rho_{B}  &  \approx12\left(  \sin^{2}\alpha\sin F\right)  d\alpha\wedge
dF\wedge dG\neq0\ . \label{prop1.21}%
\end{align}
If one takes into account Eqs. (\ref{prop1.1}), (\ref{prop1.2}) and
(\ref{prop1.21}) together with the fact that both $F$ and $G$ depend linearly
on the coordinates defined in Eqs. (\ref{Minkowski}) and (\ref{period0}), one
can see easily that Eqs. (\ref{equ1.1}) and (\ref{equ1.2}) are identically
satisfied and that Eq. (\ref{equ}) reduces to Eq. (\ref{equ2}) $P\left[
\alpha\right]  =0$. Thus, unlike what happens in the case of the original
spherical Skyrme ansatz discussed in the previous section, in the present case
one can reduce the three Skyrme field equations to only one consistent
equation for $\alpha(r)$ because the potentially dangerous terms involving
$\sin F$ appearing in Eq. (\ref{equ}) are eliminated by the
property\footnote{It is worth to note that in the usual case of the spherical
hedgehog introduced by Skyrme himself (see Eqs. (\ref{original1}) and
(\ref{original2})) $\nabla_{\mu}G\nabla^{\mu}G\neq0$.} in Eq. (\ref{prop1.2}).

All in all, the three coupled Skyrme field equations Eq.
(\ref{nonlinearsigma1}) with the ansatz in Eq. (\ref{ans1}) reduce to a simple
quadrature which can be integrated using elliptic functions. The boundary
condition in Eq. (\ref{bc1}) reduces to:%
\begin{equation}
\int_{0}^{n\pi}\frac{d\alpha}{\eta\left(  \alpha,E_{0}\right)  }=n\int
_{0}^{\pi}\frac{d\alpha}{\eta\left(  \alpha,E_{0}\right)  }=2\pi
\ ,\ E_{0}>0\ . \label{equ4}%
\end{equation}
The above equation for $E_{0}$ always has a positive real
solution\footnote{The left hand side of Eq. (\ref{equ4}) as function of
$E_{0}$ increases from very small values (when $E_{0}$ is very large and
positive) to very large values (when $E_{0}$ is close to zero but positive).
Thus, there is always a value of $E_{0}$ which satisfies Eq. (\ref{equ4}).}.
Moreover, one can see that $\partial_{r}\alpha>0$ and that, when $n$ is large,
both $\eta\left(  \alpha,E_{0}\right)  $\ and $E_{0}$ are of order $n$.

The conclusion is that the ansatz in Eqs. (\ref{standard1}),
(\ref{standard1.1}), (\ref{standard1.112}) and (\ref{ans1}) in the flat metric
in Eqs. (\ref{Minkowski}) and (\ref{period0}) in which the profile $\alpha$ is
given in closed form in Eqs. (\ref{equ2.0}), (\ref{equ3}), (\ref{equ3.1}) and
(\ref{equ4}) \textit{gives rise to exact and topologically non-trivial
solutions (with Baryonic charge }$np$\textit{) of the complete set of Skyrme
field equations}\footnote{Here we have shown this both in the matrix notation
(see Eq. (\ref{nonlinearsigma1})) and in the explicit parametrization in terms
of $\alpha$, $F$ and $G$ (see Eqs. (\ref{equ}), (\ref{equ1.1}) and
(\ref{equ1.2})).} for any integers $n$ and $p$ and for any odd integer $q$.

\subsection{Energy-density}

The energy density (replacing $\left(  \partial_{r}\alpha\right)  ^{2}$ with
$\eta\left(  \alpha,E_{0}\right)  ^{2}$\ using Eq. (\ref{equ3})) in Eq.
(\ref{timunu1}) with the ansatz in Eq. (\ref{ans1}) reads%
\begin{equation}
T_{00}=\frac{Kp}{4L^{2}A}\left[  2\rho_{0}\left(  \alpha\right)  +4\sin
^{2}\left(  q\theta\right)  \rho_{1}\left(  \alpha\right)  \right]  \ ,
\label{enden1}%
\end{equation}
where%
\begin{align}
\rho_{0}\left(  \alpha\right)   &  =\frac{\left(  Lq\right)  ^{2}}{p}\sin
^{2}\alpha+\eta\left(  \alpha,E_{0}\right)  ^{2}\left[  \frac{\left(
L\right)  ^{2}}{p}+\lambda\frac{\left(  Lq\right)  ^{2}}{Ap}\sin^{2}%
\alpha\right]  \ ,\ \label{enden2}\\
\rho_{1}\left(  \alpha\right)   &  =\sin^{2}\alpha\left[  Ap+\lambda
q^{2}p\sin^{2}\alpha+\lambda p\eta\left(  \alpha,E_{0}\right)  ^{2}\right]
\ . \label{enden3}%
\end{align}

There are \textit{two important differences} with respect to the first
analytic examples of Skyrmions living at finite density in flat spaces
\cite{Fab1} \cite{gaugsk} \cite{gaugsk2}.

\textit{Firstly}, in that references, the factor $\sin2H$ (where $H$ is the
profile defined in Eqs. (9) and (10) of \cite{Fab1}) appears linearly in the
Baryon density in Eq. (16) of \cite{Fab1}. Thus, in that references, it is not
possible to increase the Baryon charge by increasing the number of "bumps" in
the profile $H$.

\textit{Secondly}, the energy density (defined in Eq. (15) of \cite{Fab1})
only depends on $H$ (and, consequently, it only depends on one spatial
coordinate). This prevents one from describing explicitly crystal-like
structures in which the number of bumps in the energy-density is related with
the Baryon number (as, in that references, only one bump in $H$ is allowed).

In the present case both problems are solved since in the Baryon density in
Eq. (\ref{ansch}) $\sin\alpha$ appears quadratically (so that by increasing
the number of "bumps" in $\alpha$ one also increases the Baryon charge) and
the \textit{energy density in Eqs. (\ref{enden1}), (\ref{enden2}) and
(\ref{enden3}) (even after taking the trace over the} $SU(2)$ \textit{indices)
depends non-trivially both on the profile} $\alpha$ \textit{and on the spatial
coordinate} $\theta$ (so that a clear crystal-like structure in its peaks emerges).

The two contour plots of the energy-density\footnote{Note that one can replace
the $r$-dependence with the $\alpha$-dependence in the energy density using
Eq. (\ref{equ3}) to eliminate $dr$ and $\partial_{r}\alpha$ in favour of
$d\alpha$ and $\eta\left(  \alpha,E_{0}\right)  $.} in the $r-\theta$-plane
below in \textit{Figure 1} (with $p=1$, $q=1$ and $n=25$ on the left and
$p=1$, $q=3$ and $n=25$\ on the right) show the crystal-like pattern of the
bumps (whose positions can be determined explicitly maximizing the energy-density):



\begin{figure}[h]
\label{Energy density} \centering
\begin{subfigure}[t]{0.48\textwidth}
\centering
\includegraphics[width=0.9\linewidth]{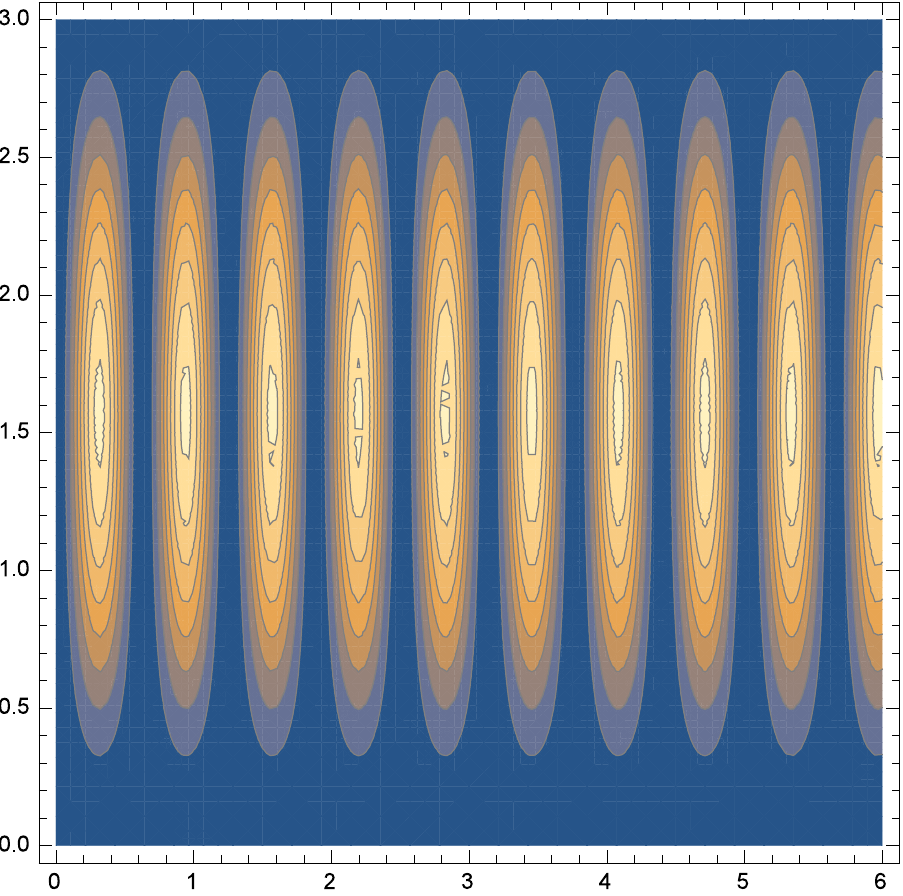}
\caption{$p=1$, $n=25$ and $q=1$ } \label{fig:f1}
\end{subfigure}
\hfill\begin{subfigure}[t]{0.48\textwidth}
\centering
\includegraphics[width=0.9\linewidth]{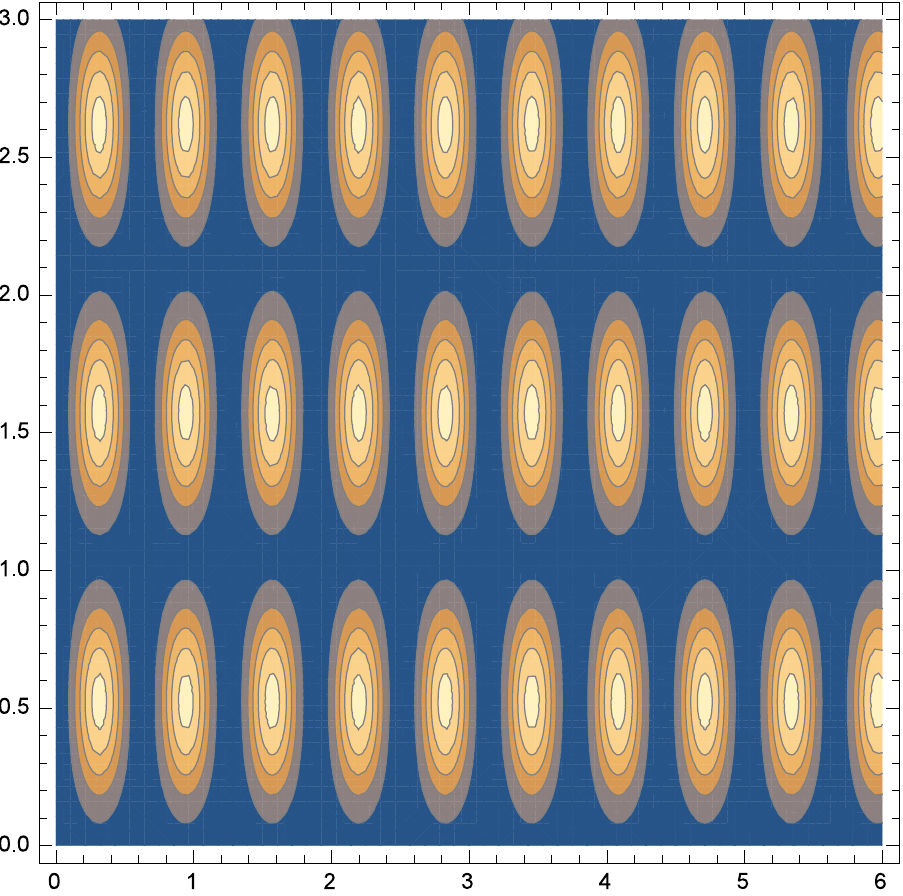}
\caption{$p=1$, $n=25$ and $q=3$} \label{fig:f2}
\end{subfigure}\caption{The left panel (a) shows the contour plot of the
energy density with $p=1$, $n=25$ and $q=1$ (where the highest energy density
is in clear yellow while the lowest is blue). The right panel (b) shows the
contour plot of the energy with $p=1$, $n=25$ and $q=3$ (in both cases it has
been assumed that $K=2$ and $\lambda=1$; note that the horizontal axis
corresponds to the $r$ direction while the vertical axis to the $\theta
$\ direction). The patterns of the peaks clearly show a crystal-like
structure. It is clear to see that, when passing from $q=1$ to higher values
of $q$ (keeping fixed $n$ and $p$), the bumps in the energy density (which,
when $q=1$, carry one unit of Baryon charge) are divided into $q$ smaller
bumps each of which carries $1/q$ of Baryon charge.}%
\end{figure}

These configurations are ordered arrays of Baryonic tubes (namely, tubes in
which most of the energy density and Baryonic charge is concentrated within
tube-shaped regions). It is easy to recognize that this is the case since both
the energy density and the Baryon density do not depend on the third spatial
direction $\phi$ \ (while they are periodic both in $r$ and in $\theta$). A
three-dimensional contour plot would reveal that the \textit{clear-yellow
regions} in \textit{Figure 1} of maximal energy density would become, in 3D,
\textit{clear-yellow} tubes. The positions of these Baryonic tubes in the
$r-\theta$ plane (defined by the clear-yellow spots in \textit{Figure 1})
manifest a clear crystalline order. The similarity with the spaghetti-like
configurations found (numerically) in the nuclear pasta phase (see the plots
in \cite{pasta0} \cite{pasta01} \cite{pasta02} \cite{pasta} and references
therein) is quite amazing.

Consequently, the interpretation of the above results together with the plots
in \textit{Figure 1} tell that, when $q=1$, the integer $n$ corresponds to the
number of Baryonic tubes in the configurations while $p$ is the Baryonic
charge of each tube. The role of the integer $q$ will be analyzed in the next section.

The total energy $E_{tot}$ of the configuration can be obtained in a closed
form (using Eqs. (\ref{equ3}) and (\ref{equ3.1})). One is left with an
integral of an explicitly known function of $\alpha$ which contains all the
relevant informations:%
\begin{align}
E_{tot} &  =\int\sqrt{-g}d^{3}xT_{00}=B\pi^{2}K\int_{0}^{\pi}d\alpha
\Omega\left(  \alpha;p,q,n\right)  \ ,\label{finalflat}\\
\Omega\left(  \alpha;p,q,n\right)   &  =\frac{A\sin^{2}\alpha}{\eta\left(
\alpha,E_{0}\right)  }\left[  \frac{Lq^{2}}{Ap}+\frac{p}{L}+\frac{\lambda
pq^{2}}{AL}\sin^{2}\alpha\right]  +\label{finalflat2}\\
&  +\frac{\eta\left(  \alpha,E_{0}\right)  }{A}\ \left[  \frac{AL}{p}%
+\lambda\sin^{2}\alpha\left(  \frac{Lq^{2}}{p}+\frac{Ap}{L}\right)  \right]
\ ,\nonumber
\end{align}
where $B=np$ is the Baryon number, $n$ is the number of bumps associated with
the profile $\alpha$ in the $r$ direction. The above equation represents the
explicit expression for the total energy as a function of the parameters of
the system. As it was already emphasized in the discussion below Eq.
(\ref{maxcond}), the total energy does not depend on $p$ and $L$ separately
but only on the ratio $p/L$. 

It is convenient from the energetic point of view neither to have very small
$p/L$ nor to have very large $p/L$. This can be seen as follows: let us denote
the variable $p/L$ as $\xi$. Then, the total energy per-Baryon (namely,
$E_{tot}/B$ from Eqs. (\ref{finalflat}) and (\ref{finalflat2})) as function of
$\xi$ (assuming that $n$ is large so that $\eta\left(  \alpha,E_{0}\right)
\sim n$) reads%
\begin{align}
\frac{E_{tot}\left(  \xi\right)  }{B}  & =c_{0}\xi+\frac{c_{1}}{\xi
}\ ,\ \ \ \xi=\frac{p}{L}\ ,\label{totencsi}\\
c_{0}  & >0\ ,\ c_{1}>0\ ,\label{totencsi2}%
\end{align}
where $c_{1}$ and $c_{0}$ (which can be read from the explicit expressions in
Eqs. (\ref{finalflat}) and (\ref{finalflat2})) depend on all the other
parameters of the theory. Thus, the total energy per-Baryon as function of
$\ \xi$ has a non-trivial local minimum:%
\begin{equation}
\xi^{\ast}=\sqrt{\frac{c_{1}}{c_{0}}}\ .\label{totencsi3}%
\end{equation}
This fact means that if one increases the value of $p$ one also has to
increase the length of the tube $L$ in such a way to keep $p/L$ equal to its
optimal value in the above equation. Therefore one can interpret $p$ as the
length of the Baryonic tubes in unit of $fm$.

A further natural question is the following:

\textit{given a total Baryon number }$B$\textit{, is it energetically more
convenient to have higher }$p$\textit{ and lower }$n$\textit{ or the other way
around}?

The answer to this question in the low energy limit of QCD can be found just
by analyzing the function $\Omega\left(  \alpha;p,q,n\right)  $\ defined in
Eq. (\ref{finalflat2}) which contains all the relevant informations. For
instance, can see explicitly that if $A$ decreases it is better to have larger
values for $p$. 

In other words, if one keeps $B=np$ fixed and $A$ decreases, then it is better
to have taller tubes (and a smaller number of tubes to keep $B$ fixed).

On the other hand, if one increases $A$ it is better to have a bigger number
of smaller tubes (to keep $B$ fixed). Although a plot of the total energy (or
total energy per-Baryon as function of all the parameters is not very useful),
if one keeps some of the parameters fixed and only considers pairs of related
parameters (such as $p$ and $L$ or $A$ and $n$) the pattern is always similar
to the one in Eqs. (\ref{totencsi}) and (\ref{totencsi2}). I hope to come back
on the properties of the function $\Omega\left(  \alpha;p,q,n\right)
$\ defined in Eq. (\ref{finalflat2}) in a future publication.

\subsection{A remark on the stability}

A remark on the stability of the above crystals is in order. In many
situations, when the hedgehog property holds (so that the field equations
reduce to a single equation for the profile) the most dangerous perturbations
are perturbations of the profile which keep the structure of the ansatz (see
\cite{shifman1} \cite{shifman2} and references therein). In the present case
these are%
\begin{equation}
\alpha\rightarrow\alpha+\varepsilon u\left(  r\right)  \ ,\ \ \ \varepsilon
\ll1\ . \label{pert}%
\end{equation}
It is a direct computation to show that the linearized version of Eq.
(\ref{equ2}) around a background solution $\alpha_{0}\left(  r\right)  $ of
charge $B=np$ always has the following zero-mode: $u\left(  r\right)
=\partial_{r}\alpha_{0}\left(  r\right)  $. Due to Eqs. (\ref{equ3}),
(\ref{equ3.1}) and (\ref{equ4}) $u(r)$ has no node so that it must be the
perturbation with lowest energy. Thus, the present solutions are stable under
the above perturbations. It is also worth to remark that isospin modes
$U^{A}=A^{-1}U_{0}A$ (where $U_{0}$ is the solution of interest and $A$ is a
generic $SU(2)$ matrix which only depends on time) have positive energies (if
the energy of $U_{0}$ is positive, as in the present case) \cite{ANW}. The
effective action for these modes is the one of a spinning top and the energy
of the corresponding perturbations is positive definite. Isospin modes are
"transverse" to the perturbations of the profile in Eq. (\ref{pert}) as they
do not touch $\alpha(r)$: the reason is that both $U^{A}=A^{-1}U_{0}A$ and
$U_{0}$ have the same profile. This stability argument holds for any $n$, $p$
and $q$ in Eqs. (\ref{standard1}), (\ref{standard1.1}), (\ref{ans1}) and
(\ref{bc1}).

It is worth to note that in the original spherical ansatz of Skyrme with
radial profile $\alpha_{s}$ discussed in the previous sections, the
topological density is also proportional to $\sin^{2}\alpha_{s}$ so that one
can increase the winding increasing the "spherical bumps" in the energy
density (which in the case of the original spherical ansatz of Skyrme only
depends on the radius). As it is well known, one can construct (numerically)
these "higher charges spherical Skyrmions" but all of them are unstable. The
instability arises from zero modes of the form $\partial_{R}\alpha_{s}$ with
nodes (so that there are perturbations with negative energies and, indeed, the
only stable solution of this family is the spherical Skyrmion of charge 1).

\section{Bumps with fractional Baryonic charge}

The role of the odd-integer $q$ (which does not enter directly in the Baryon
charge in Eq. (\ref{bc1})) has not been discussed. A clear hint is the
comparison of the energy-density contour plots above for two such crystal-like
structures with the same Baryon charges, $p$ and $n$ but different $q$ (for
instance, $p=1$, $q=1$ and $q=3$). In these plots, there are $n$ bumps in the
first structure with $q=1$ which correspond exactly to the Baryon charge of
the layers: thus, each bump carries one unit Baryon charge as expected. In the
second structure with $q=3$ there are $3n$ bumps in the $r-\theta$ plane but
the Baryon charge has not changed. Thus, each bump carries $1/3$ (in general
$1/q$) of the unit topological charge. Thus, the integer $q$ is related to the
fractional Baryonic charge carried by the bumps of the crystal. In the generic
case in which $p\neq1$ the interpretation is the following. The number of
Baryonic tubes $N_{tubes}$ in the solution is $N_{tubes}=qn$. On the other
hand, each tubes carries a Baryonic charge equal to $\frac{p}{q}$ so that the
total Baryonic charge is still $B=np$.

\subsection{An example of topologically trivial solution}

Here it is useful to discuss the following natural question:

\textit{Since both the energy-density and the Baryon density only depend on
two spatial coordinates, are the solutions constructed above really
three-dimensional?}

It is possible to answer to the above question just using the definition:
since the above configurations in Eqs. (\ref{standard1}), (\ref{standard1.1}),
(\ref{standard1.112}) and (\ref{ans1}) in the flat metric in Eqs.
(\ref{Minkowski}) and (\ref{period0}) in which the profile $\alpha$ is given
in closed form in Eqs. (\ref{equ2.0}), (\ref{equ3}), (\ref{equ3.1}) and
(\ref{equ4}) solve the complete set of Skyrme field equations\footnote{As
shown both in the matrix form in Eq. (\ref{nonlinearsigma1}) and in the
explicit parametrization in Eqs. (\ref{equ}), (\ref{equ1.1}) and
(\ref{equ1.2}).
\par
{}} in (3+1)-dimensions and, moreover, the topological charge is
non-vanishing, then according to the classic references \cite{skyrme}
\cite{bala0} \cite{witten0} \cite{ANW} these configurations are genuine
three-dimensional objects.

In fact, it is more useful to give an example of a configurations which looks
similar to the ones analyzed in the previous section but, in fact, has
vanishing topological density.

A "topologically trivial" ansatz for the Skyrme field is
\begin{equation}
\alpha=\alpha\left(  r\right)  \ ,\ F=q\theta\ ,\ G=0\ ,\ q\in%
\mathbb{N}
\ \ .\label{trivial}%
\end{equation}
It is easy to see that the topological density vanishes (since $G=0$). 

On the other hand, the above ansatz still gives rise to non-trivial Skyrme
field equations for $\alpha$ and $F$. In the explicit parametrization in Eqs.
(\ref{equ}), (\ref{equ1.1}) and (\ref{equ1.2}) one can see that, with the
choice in Eq. (\ref{trivial}), Eqs. (\ref{equ1.1}) and (\ref{equ1.2}) are
identically satisfied while Eq. (\ref{equ}) reduces (once again) to Eq.
(\ref{equ2}). Despite the fact that both in this trivial case and in the
topologically non-trivial configurations analyzed previously the equations for
the profile $\alpha$ are the same, the corresponding configurations are
totally different. 

One obvious difference is that there is no topological argument which prevents
the configuration defined in Eq. (\ref{trivial}) from decaying into the
trivial vacuum as the corresponding Baryon charge vanishes. Another difference
is that the total energy of the trivial configuration defined in Eq.
(\ref{trivial}) which reads (assuming the same boundary conditions for
$\alpha$ and $F$ so that the definition of $\eta\left(  \alpha,E_{0}\right)
$\ is the same as in the previous sections):%

\begin{align}
E_{trivial} &  =\int\sqrt{-g}d^{3}xT_{00}^{trivial}=n\pi^{2}K\int_{0}^{\pi
}d\alpha\Omega^{trivial}\left(  \alpha;q,n\right)  \ ,\label{trivial2}\\
\Omega^{trivial}\left(  \alpha;q,n\right)   &  =\frac{\sin^{2}\alpha}%
{\eta\left(  \alpha,E_{0}\right)  }Lq^{2}+\frac{\eta\left(  \alpha
,E_{0}\right)  }{A}\ \left[  AL+\lambda\sin^{2}\alpha\left(  Lq^{2}\right)
\right]  \ .\label{trivial3}%
\end{align}
The differences between this expression and the ones in Eqs. (\ref{finalflat})
and (\ref{finalflat2}) can be ascribed to the genuine three-dimensional nature
of the Baryonic tubes discussed in the previous sections and to the effective
two-dimensional nature of the topologically trivial configurations defined in
Eq. (\ref{trivial}). In particular, comparing the expressions in Eqs.
(\ref{finalflat}) and (\ref{finalflat2}) with Eqs. (\ref{trivial2}) and
(\ref{trivial3}), one can observe that to have Baryonic charge is quite
expensive from the energetic point of view.

The example in this subsection discloses neatly the differences between
configurations with and without Baryonic charge.

\section{Conclusions}

Analytic topologically non-trivial solutions of the complete set of Skyrme
field equations of charge $np$ describing ordered arrays of Baryonic tubes
with crystalline structure living at finite density have been constructed
explicitly. These configurations are characterized by three integers: $n$, $p$
and $q$ ($q$ \ being odd). When $q=1$, the Baryonic charge of each tube is $p$
while $n$ is the number of tubes so that the total Baryonic charge is $B=np$.
On the other hand, when $q$ is greater than $1$, the Baryon charge of each
tube is $p/q$ while the total number of tubes is $nq$ (so that the total
Baryonic charge is still $B=np$). The positions of the peaks in the
energy-density can be computed explicitly by maximizing the energy density.
These configurations pass a non-trivial stability test.

\subsection*{Acknowledgements}

This work has been funded by the Fondecyt grants 1160137. The author would
like to thank C. Martinez and anonymous referees for useful suggestions. The
Centro de Estudios Cient\'{\i}ficos (CECs) is funded by the Chilean Government
through the Centers of Excellence Base Financing Program of Conicyt.

\end{document}